\documentclass[useAMS,usenatbib]{mn2e}

\usepackage{graphicx}
\usepackage{placeins}
\usepackage{amsfonts,amsmath,amssymb}
\usepackage{url}
\usepackage{enumerate}
\usepackage{rotating}
\usepackage{hyperref}
\usepackage{url} 

\usepackage{float,booktabs,multirow}
\usepackage[caption=false]{subfig}

\usepackage[usenames]{color}

\title[Zeldovich Pancakes]{Zeldovich pancakes at redshift zero: the
  equilibration state and phase space properties}

\author[{Wadekar} \& {Hansen}]
{Digvijay Wadekar$^1$ \& Steen H. Hansen$^2$\\
$^1$Department of Physics, Indian Institute of Technology Bombay, Mumbai\\
$^2$Dark Cosmology Centre, Niels Bohr Institute, University of Copenhagen,
Juliane Maries Vej 30, 2100 Copenhagen, Denmark\\
digvijay.wadekar93@iitb.ac.in, hansen@dark-cosmology.dk}

\begin{document}

\date{\today}

\pagerange{\pageref{firstpage}--\pageref{lastpage}} \pubyear{2013}

\maketitle

\label{firstpage}

\begin{abstract}
  One of the components of the cosmic web are sheets, which are
  commonly referred to as Zeldovich pancakes. These are structures
  which have only collapsed along one dimension, as opposed to
  filaments or galaxies and cluster, which have collapsed along two or
  three dimensions.  These pancakes have recently received renewed
  interest, since they have been shown to be useful tools for an
  independent method to determine galaxy cluster masses.

  We consider sheet-like structures resulting from cosmological
  simulations, which were previously used to establish the cluster
  mass determination method, and we show through their level of
  equilibration, that these structures have indeed only collapsed along
  the one dimension. We also extract the density profiles of these
  pancake, which agrees acceptably well with theoretical expectations.
  We derive the observable velocity distribution function (VDF)
  analytically by generalizing the Eddington method to one dimension,
  and we compare with the distribution function from the numerical
  simulation.
\end{abstract}


\graphicspath{{Images/}}
\maketitle

\section{Introduction}

The large-scale structure of the universe as observed in redshift
surveys displays a complicated geometry.  This complex structure
referred to as the "cosmic web", is characterized by the presence of
clusters, filaments, sheets and voids. More than 30 years ago galaxy
surveys found evidence of filaments \citep{1981ApJ...243..411G}
and analyses of modern surveys find hundreds of filaments and
pancakes \citep{costa-duarte2011}.  Numerical simulations complement
the obserations, and find that sheets (which include walls and
pancakes) may occupy over half of the volume \citep{Hahn2007a}.

The structure of the cosmic web is the result of the gravitational
growth of the small amplitude primordial density and velocity
perturbations.  \citet{1970A&A.....5...84Z} first suggested the
anisotropic nature of gravitational collapse leading to Zeldovich
Pancakes, which are formed when an overdense region collapses first
along one of its principal axes. The pancake picture was further
developed in \citet{1982GApFD..20..111A} and \citet{shandarin1989}.
Various papers have simulated Zeldovich pancakes under controlled
conditions
\citep{1984ApJ...281....1F,1998ApJ...509...62T,Binney-2004,Schulz-2013}
which typically involve simulations of particles only along a fixed
axis perpendicular to the pancake.

The dynamics of pancakes has been studied extensively with large
numerical $N-$body simulations, where pancakes have been observed over
the last almost 20 years \citep{1995PhRvL..75....7S}. The dynamical
importance of these sheets, expressed through the mass fraction of
sheets in the cosmic web is rather uncertain, and varies in the range
$(6\% - 36\%)$, while the volume fraction varies from $(5\% - 54\%)$ 
\citep{cautun2014,forero-romero-2009,hoffman2012,Hahn2007a,Shandarin2012,
2010MNRAS.408.2163A,Shen06,Doroshkevich70}.

There are several ways of defining whether a structure in the cosmic
web is a filament, sheet, void or halo.  Often, structures are
differentiated on the basis of eigenvalues of the second derivative of
the tidal tensor $T_{ij}$, for a sheet $T_{ij}$ has only one positive
eigenvalue. Alternatively the sheets can be defined through the shear
tensor, which is a rescaled derivative of the velocity field.
\citep{2007MNRAS.381...41H, forero-romero-2009,
  2010MNRAS.409..156B,costa-duarte2011,hoffman2012}.  In recent
studies, the entire 6 dimensional phase space information is taken into
account for recognizing structures \citep{Shandarin2012, abel2012,
Neyrinck12, Neyrinck2012b}.

Direct observations of pancakes is extremely difficult whereas
identifying halos is straight forward. This is due to their planar nature
and their low contrast with the background \citep{Aragon07,
2010MNRAS.408.2163A, cautun2014}.  Generally,
there is a non uniform mass distribution along the area of the sheet,
since most of the sheet mass is located in a few very massive regions
whereas many of the remaining regions have very low density with no
massive halos in them.  Observations of high-density regions in
galaxy observations were pioneered 60 years ago
\citep{devaucouleurs53,abell58}, and have today led to the detection
of hundreds of sheets \citep{costa-duarte2011}.  All these sheets are,
however, based on overdensities in phase-space only, and it has still
not been clearly established to which extent the sheets have infact
collapsed along the one dimension only.  This is one of the
concerns of the present paper, to consider whether such overdensities
in phase-space are indeed sheets.

Recently it was shown, that certain pancakes may be useful as tools to measure
the mass of massive galaxy clusters \citep{Falco_Paper}. These
pancakes are positioned at 3-10 virial radii from the cluster. The cluster mass
is measured by considering the small perturbation to the Hubble flow,
which the galaxy cluster will induce on the observable line of sight
velocity of the galaxies in the pancake. One thing missing in this
analysis is, however, to confirm that the structures used are indeed
pancakes. In this paper we will confirm that this is indeed the case.

Another aspect of pancakes are their density profile perpendicular to
the pancake.  In the papers \citep{Binney-2004, Schulz-2013} it was
shown that cold collapse leads to final equilibria that are close to a
universal power-law density profile, with an exponent less than 1/2.
Here we have compared the power law density profile for the sheet
observed in a cosmological simulation, and we find that the
theoretical prediction are in fair agreement with the result from the
cosmological simulation, however, with a slightly more shallow central
profile.

Observationally, when a pancake is seen face-on, it is not possible to
extract the density profile. However, we derive an equivalent to the
Eddington method \citep{Eddington1916, BinneyBook} in one
dimension, which in principle may allow one to extract information
about the line of sight density profile, by using the observable
distribution of velocities.

\section{Numerical Simulation of Sheets}
\label{Numerical_Simulation}

In order to consider the equilibration level of a pancake, we need to
extract a real pancake from a cosmological simulation.  A series of
recent papers \citep{Shandarin2012, abel2012, Vogelsberger:2008, White:2009, Vogelsberger:2011} developed
powerful cosmological simulation techniques to analyze the
evolution of dark matter sheet.
In this paper, we have analysed
  the sheet at a fixed redshift, $z=0$.  We consider a few of the pancakes used
in \citet{Falco_Paper}, which are based on the N-body simulation
described there (for more details, see
\citep{kravtsov97,gottloeber2008}). 
These pancakes were used to demonstrate a new method
of measuring the cluster mass, but it was not quantified if these were
truly  sheets (equilibrated along only one dimension) or instead 
filaments (equilibrated along two dimensions).

The reason for considering these particular pancakes is that they have
been found through a detection method, which will work on directly
observable data.  As observational quantities one uses only the
two-dimensional position on the sky and the line of sight
velocity. Here we will, however, analyze these structures using all the
information available from the numerical simulation.

The pancakes are pointing almost radially from a large galaxy cluster,
and lie at a distance of the order 10-40 Mpc from the cluster. 
The pancakes are extracted by considering a
slice on the sky ($\pi/4$), and then looking for overdensities in the
phase-space of projected radius and line of sight velocity, ($R,
v_{\rm los}$).  A sheet produces an almost straight line in that
space.

\begin{figure}
\includegraphics[scale=0.7,keepaspectratio=true]{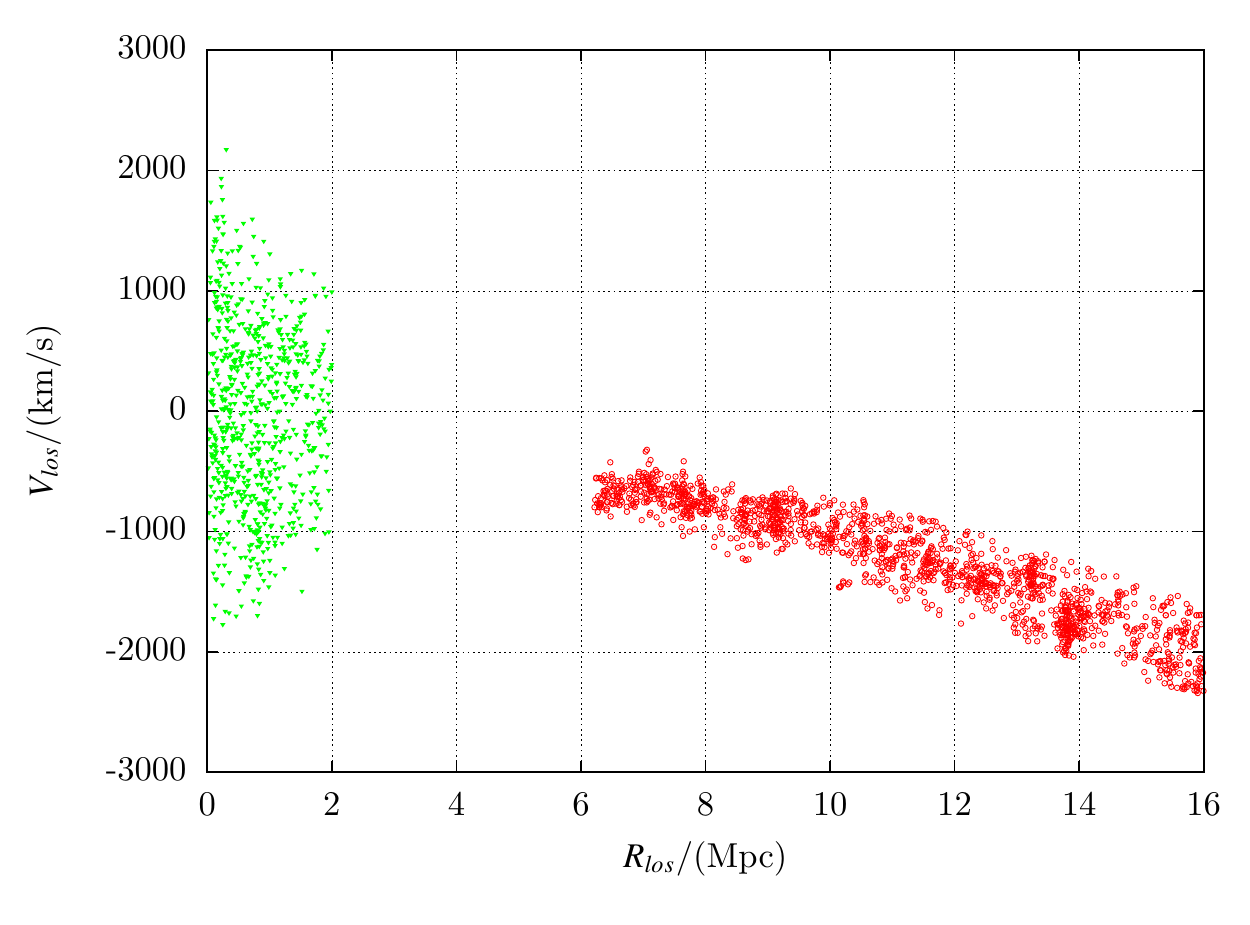}   
\caption{Phase space plot for the extracted sheet. The green points
at projected radii smaller than 2 Mpc belong to the central cluster, and
the dense, narrow strip between 6 and 16 Mpc is the sheet.}
\label{fig:1b.every5}
\end{figure}

In Figure~\ref{fig:1b.every5} we show the phase-space plot for the cluster
and the extracted sheet. The cluster is a (normal) dense collection of galaxies
at projected radii smaller than 2 Mpc, whereas the sheet forms a dense, and
fairly narrow strip at projected distances between 6 and 16 Mpc. The
velocities are centered at the cluster, and the sheet particles have
projected velocities between 500 and 2500 km/sec. Only every 5th point is
plotted.

Now, having these sheets, we can rotate them according to their major
axes (X,Y,Z), in such a way that X is an axis along the longest axis
of the sheet, and Z is an axis along the shortest axis. The latter
axis, Z, thus corresponds to the direction perpendicular to the sheet,
namely the direction where we anticipate that the collapse has already
happened.

\begin{figure}
\includegraphics[scale=1.,keepaspectratio=true]{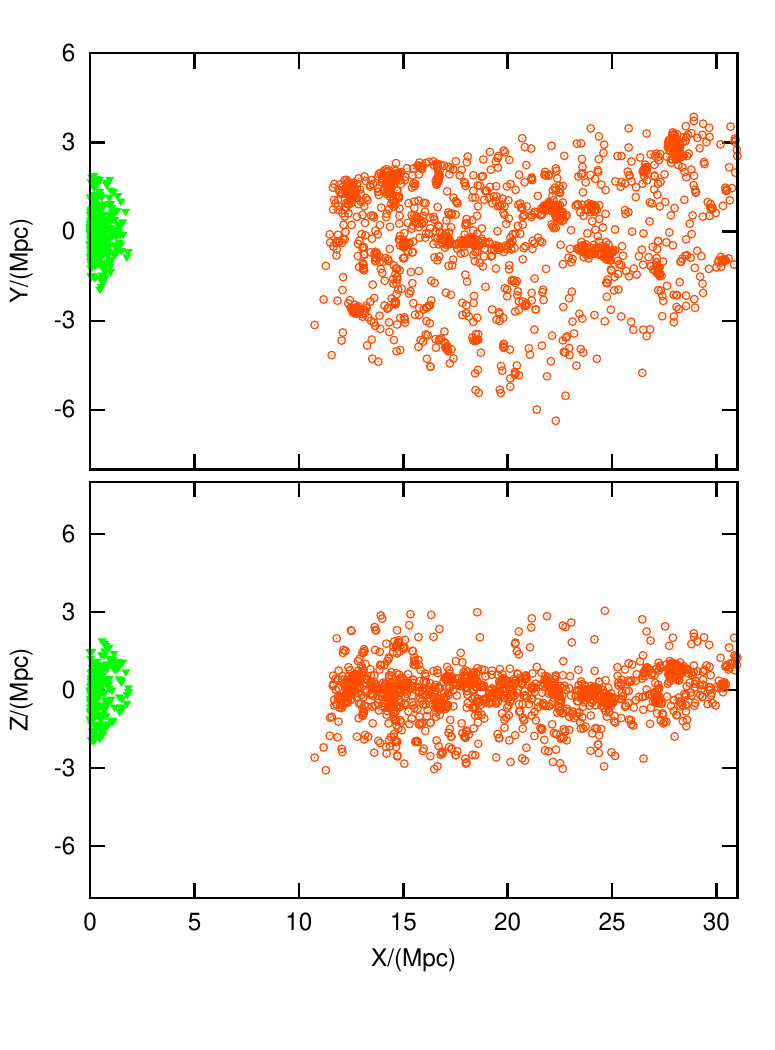}   
\caption{Real space projection of the extracted and rotated sheet.
  The X-axis corresponds to the distance from the cluster, which is
  the green points at radius below a few Mpc. The sheet are the red
  points at distances between 10 and 30 Mpc.  We clearly see visually
  that the one dimension (Z, lower panel) is closer to equilibration
  than the other dimension (Y, upper panel).}
\label{fig:2b.every5}
\end{figure}

This is demonstrated in Figure~\ref{fig:2b.every5} where we show
the real-space projection of the sheet. The x-axis is the distance
from the cluster, and we see that the sheet sits in real space
at a distance between 10 and 30 Mpc from the cluster. Visually we
clearly see, that the sheet is very uniform and broad in the 
non-collapsed dimension (Y, upper panel), whereas it is much more
compact and dense in the collapsed dimension (Z, lower panel).

As a first test, we compare the density profiles along these
directions.  We confirm that the profile along the Z-direction does
appear fairly ``\emph{gaussian}'', whereas the profiles along the
other directions (which are expected to not have collapsed yet) are
still fairly top-hat shaped, in agreement with the similar findings of
\citep{boys}.

We also need a selection of galaxies to compare with, a background.
For that we consider the same slice on the sky as the signal, however,
for the velocity we extract the region with the opposite sign. This
means that if the sheet is positioned beyond the cluster (with
positive line of sight velocity), then the background is selected as
the symmetric region, but in front of the cluster (with negative line
of sight velocities).

The largest sheet is not a homogeneous box, so we divide it into 10
radial bins (each bin containing equal number of particles) along the
X (radial) direction, and the bins are numbered in ascending order of
their distance from the cluster.  We discard the extreme bins (1 and
10).  This may in particular allow us to consider different levels of
virialization as a function of distance from the cluster.  Similar
binning is performed in the background region. 
We have analysed another two sheets from the numerical simulation.
Their properties are qualitatively similar to the sheet discussed
in this paper, however, because of a lower number of particles in those
sheets, no additional interesting lessons were learned from them.

\begin{figure*}
\includegraphics[scale=1.0,keepaspectratio=true]{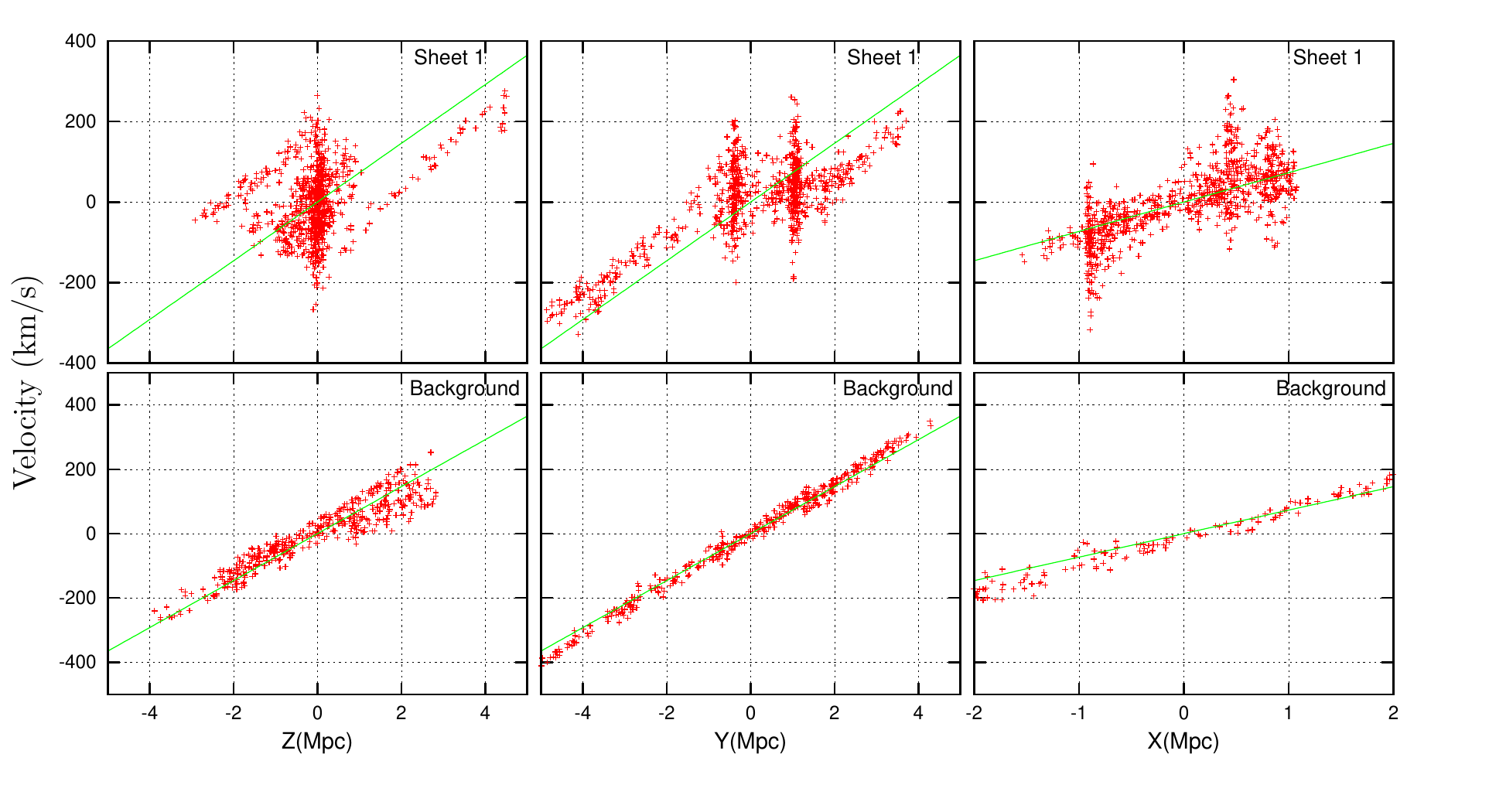}   

\caption{A sheet has been extracted using only observational data, and
here we analyse the full phase-space properties of that sheet.
The phase space plots for the 7th radial bin of the
  largest sheet in X,Y,Z directions. The origin of each plot is
  separately chosen such that mean coordinate and mean velocity of the
  particular direction are both zero.  Also, the Phase Space plots for
  the background region are provided for comparison. 
The difference in levels of
  equilibration along various axes of the sheet is clearly seen.  
The green line shown in the diagrams represents the
  Hubble Flow.}

\label{fig:Phase_Space_Complete}

\end{figure*}

\section{Stability analysis using Phase Space Plots}
\label{Phase_Space_Stability}

\subsection{Equilibration along different axes}

The structure of phase space along different axes can be used to
understand the level of equilibration along the particular axes of the
structure.  For a structure which has reached equilibrium in all the
three directions, a phase space plot in any direction will show a
semi-gaussian blob.  On the contrary, for a structure which remains
non-equilibrated along a particular direction, the Hubble flow would
completely dominate over the gravitational attraction, and its phase
space plot along that direction will show a straight line
corresponding to the Hubble expansion.

Phase space plots for one of the radial bins of the largest
sheet are shown in figure (\ref{fig:Phase_Space_Complete}) along with
the phase space plots for the background region along the three axes.
First of all, the background region (lower panels) is clearly seen to
follow the Hubble expansion fairly accurately along all 3 directions.
Instead, the signal region (upper panels) clearly shows variation from
almost following the Hubble flow (X-direction, right-most panel) to
almost being completely equilibrated (Z-direction, left-most
panel). The Y-direction is in the early stage of equilibration, only
starting to depart from the Hubble flow.

We therefore see, that the sheet candidate indeed is a real 2
dimensional sheet, and therefore differs from a filament, which should
be collapsed along 2 dimensions.

A similar analysis for the other radial bins indicate, that the outer
bins (6-9) are all in a state which have collapsed along only one
dimension, whereas the innermost bins (2-5) are in a more advance
state, where either the equilibration has happened along 2 dimensions
for some bins, or the presence of substructure is more pronounced.
This effect can also be observed by looking at the surface mass
density values for the outermost bins(6-9) being (0.2625 to 0.39375)
$M_{\sun}/pc^2$ which is lower than the density of innermost bins(2-5)
which is (0.4375 to 0.875) $M_{\sun}/pc^2$.  We will therefore limit
our analysis to the outermost bins (6-9) for the rest of this paper.

\subsection{Phase space structure}
\label{Phase_space_structure}

\begin{figure}
\includegraphics[scale=0.7,keepaspectratio=true]{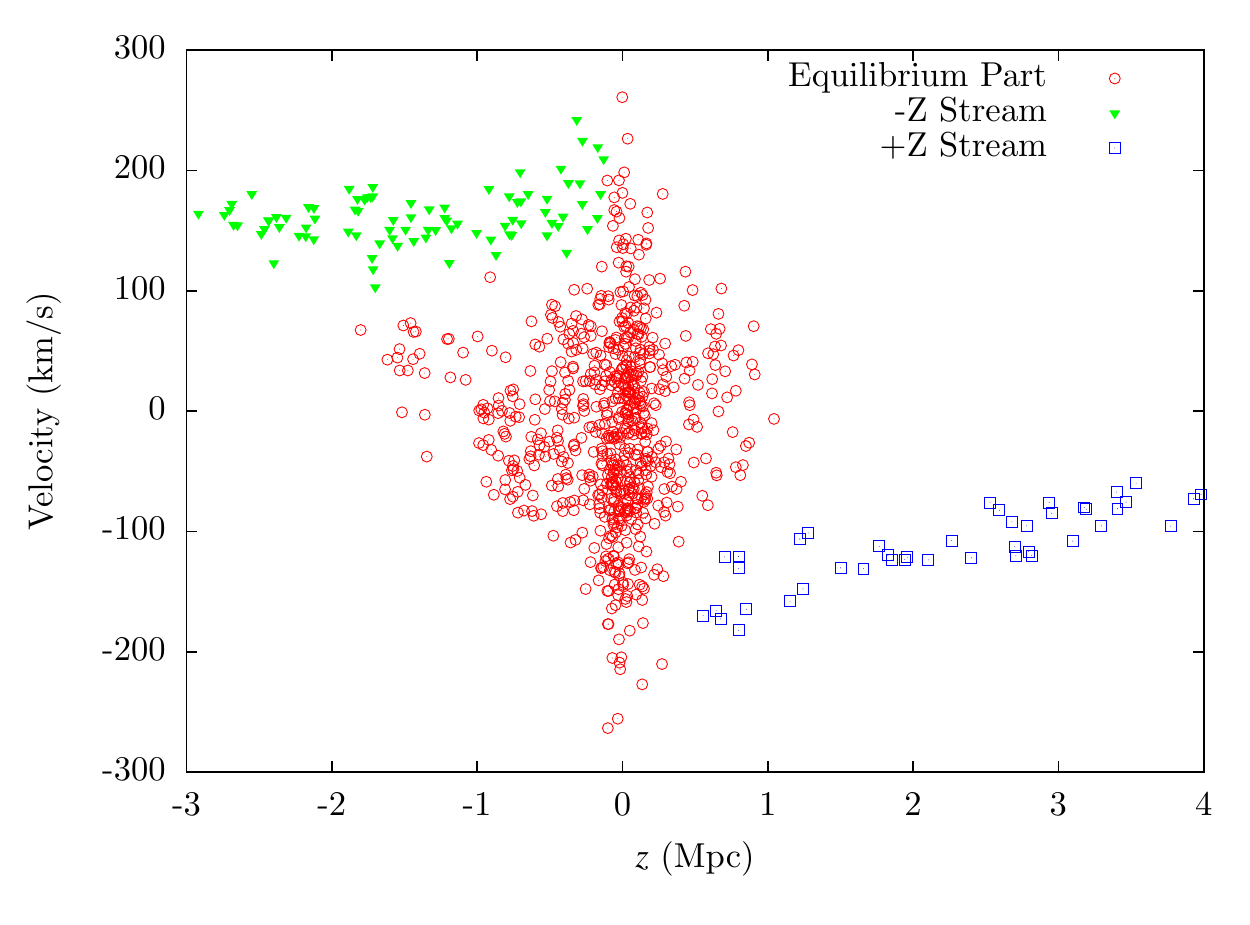}   
\caption{When we remove the Hubble flow, the phase space plot of one
  radial bin (bin 7) is seen to be divided into an almost equilibrated
  elliptical part, and 2 infalling streams of galaxies.}
\label{fig:Bin7_Parts}
\end{figure}

In this section we will consider some additional aspects of the phase
space plots along the X,Y,Z directions.  Considering their dynamic
behaviour, the simple Hubble motion of the galaxies is perturbed by
the gravitational effect of the cluster mass and the gravitational
field of the sheet itself. Thus the total velocity of each paticle
(galaxy) is the combination of two terms
\begin{equation}
\bar{v}_{r} (r) = \textup{H}\bar{r} + \bar{v}_{p} (r) \, ,
\label{eqn:Velocity_Components}
\end{equation}
which are the pure Hubble flow and an infall term $v_{p}$.  We remind
that we have rotated the sheet in such a way, that the X-direction
points radially out from the cluster.  Considering the motion along
the radial X-direction, the corresponding infall term $v_{p} (x)$ is
governed by the gravitational force of the massive nearby cluster
(which naturally acts only in the X-direction), and the contribution
from the mass of the sheet can be neglected as a first
approximation. Outside the region where the Hubble flow starts to
dominate, namely outside 3-4 times the virial radius
\citep{prada,custa}, the infall term can be approximated to be
\begin{equation}
 \overline{v}_{p} (r) \approx -v_{0} \Big(\frac{r}{r_{v}}\Big)^{-b} \, ,
\end{equation}
where $r_{v}$ is the virial radius and $v_{0}$ is an appropriate
constant (\citet{Falco_Paper}).  Thus, $\overline{v}_{p} (r)$ will
give the slight deviation from the linear Hubble flow in the slope of
the phase space plots in the X-direction for both the sheet and the
background regions, seen in the right-most panels in figure
\ref{fig:Phase_Space_Complete}. For the Y-direction, the sheet has
just started to equilibrate, so, deviation caused by the mass of the
sheet itself will be small compared to the Hubble flow, as seen in the
middle panel for the sheet.

If we subtract the Hubble flow from the velocity of each particle we
obtain the phase space figure (\ref{fig:Bin7_Parts}) for the radial bin 7.
This phase-space plot of the perpendicular direction shows the
essentially equilibrated central part, and 2 streams of galaxies in
the process of falling into the sheet.  Each radial bin in the sheet 
seems to have collapsed into a
nearly equilibrated elliptical part and two streams of infalling
galaxies which are not part of the virialized structure.
We emphasize that only directly observational data was used to
identify the sheets (2 position in projected space, and the 
line of sight velocity), whereas we now, subsequently, consider the
full detail of all of phase-space.

\begin{figure}
\includegraphics[scale=0.7,keepaspectratio=true]{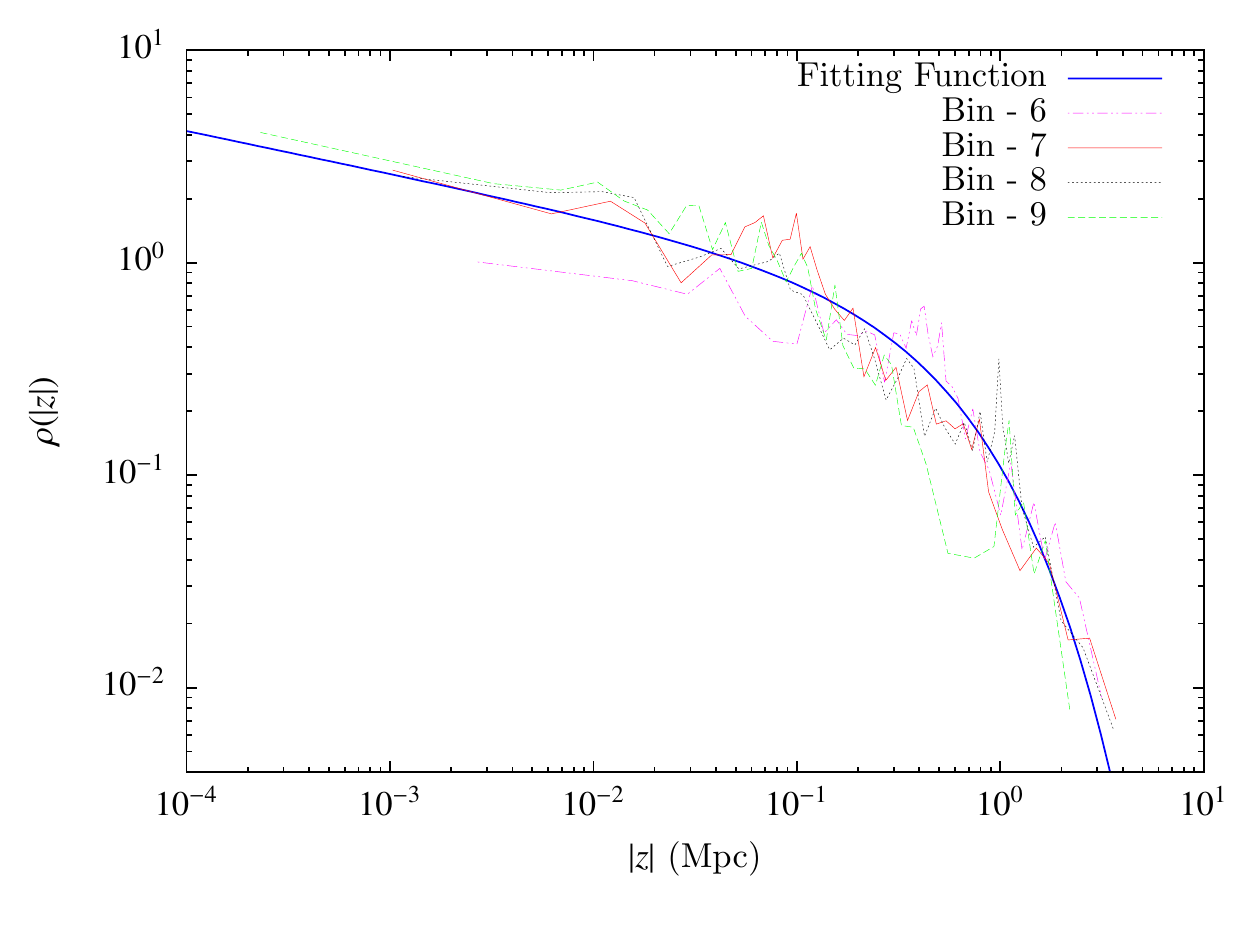}  
\caption{The density profile $\rho \left (| z |\right )$ is plotted
  for the radial bins numbered 6, 7, 8, 9 on a logarithmic scale. The blue
  fitting curve shows the model in equation
  (\ref{eqn:fitting_function}) with the value of $\gamma = 0.2$}
\label{fig:Tremaine_Comparison}
\end{figure}

By removing the particles in the two streams, we can get the density
profile of the equilibrated structure as a function of distance from
the center.  The profiles are somewhat noisy due to substructures
which are not fully equilibrated.  The structure of the equilibrated
elliptical part in the plots is similar to the plots derived from one
dimensional simulations in the paper \citep{Schulz-2013}.  This is
shown in figure (\ref{fig:Tremaine_Comparison}). We have also plotted
a smooth function given by
\begin{equation}
\rho_{z} \left ( z \right ) \sim \left | z \right |^{-\gamma } \exp\bigg(-\Big(\frac{|z|}{z_{0}}\Big)^{1-\gamma }\bigg) \, ,
 \label{eqn:fitting_function}
\end{equation}
as a fit to the curve.  From the fitting curve we find that $\gamma
= 0.2$. This can be compared to various analytical predictions:
$\gamma = 0.47$ in the paper \citep{Schulz-2013} and $\gamma = 0.5$ in
\citet{Binney-2004}.

\subsection{Behavior of Infalling Streams}

The streams in figure (\ref{fig:Bin7_Parts}) can be compared to the
stages of early equilibration as described in the papers
\citet{Binney-2004,Schulz-2013}.  The potential energy
of the sheet can be calculated from the density profile using equation
(\ref{eqn:Poisson_Equation}).
To make a better estimation of the potential energy profile,
at distances away from its center, we have taken into account the
approximations to be made for a \emph{finite} sheet.

\begin{figure}
\includegraphics[scale=0.7,keepaspectratio=true]{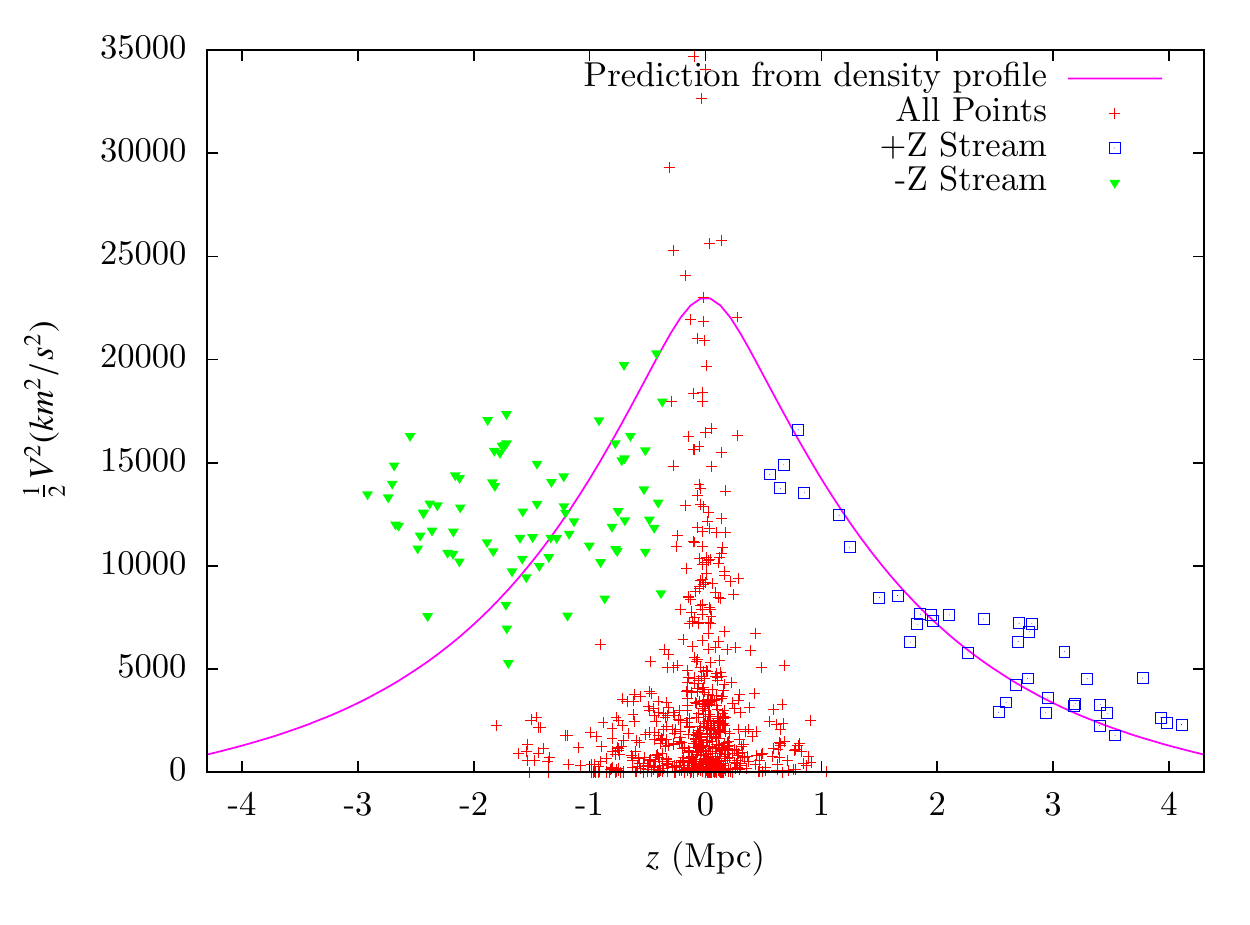}  
\caption{The kinetic energy distribution function $1/2 \,
  v_{z}^{2}\left ( z \right )$ of bin 7 are plotted.  The fitting
  function has been derived from the density profile of this bin.}
\label{fig:Energy Fit}
\end{figure}

Thus, we have estimated the total energy ($K.E + P.E$) of the
infalling particles by comparing the change in kinetic energy of the
particles in the 2 streams with the profile of potential energy due to
the sheet itself.  Energy conservation holds as long as the potential
is stable in time.  Such a constant total energy curve is given in
figure (\ref{fig:Energy Fit}). The kinetic energy of particles in both
in +Z and -Z streams are compared with the constant total energy
curve.

The galaxies in the +Z stream follow this prediction very accurately.
The stream on the other side of the sheet clearly has a component of
particles (galaxies) which do not follow this simple behaviour. These
particles are at distances 1-3 Mpc from the sheet, and have an
additional energy of approximately 150 km/sec. In figure 4 we see that
these are more dense particles, with a non-zero velocity dispersion.
A potential origin of this discrepancy may be related to the
non-equilibrated substructures, which are mainly seen in the -Z
direction, implying that the -Z stream galaxies could be in a peculiar
stage in the equilibration process rather than in a completely free
fall motion.

\section{Eddington Model for one dimension}
\label{1D Eddington Model}

In this section, we derive an equivalent of the Eddington
model \citep{Eddington1916, BinneyBook} for its application on
isotropic sheets with an infinite area.

We shall consider self-consistent systems in which the density
distribution determines the potential through Poisson’s equation, and
the potential must also determine the density consistently through the
collisionless Boltzmann equation.

As we will see, an equivalent of the Eddington
model for 3 dimensions can be used approximately for sheets along their
perpendicular direction (Z- axis).  These sheets are assumed to have
uniform spatial distribution along the X-Y plane and to be completely
equilibrated along the Z-axis.  The change in gravitational potential
or the total energy of galaxies along X and Y axis is considered to be
negligible as compared to their change along Z axis.  Thus we consider
only the Z axis values of potential and total energy of the galaxies
for our analysis.  We define the relative potential $\Psi $, and the
relative energy $\xi$ of a galaxy by 
\begin{equation}
 \Psi = -\Phi  + \Phi _{0}  \hspace{5 mm} {\rm and} \hspace{5 mm} \xi = \Psi - \frac{\upsilon^{2} }{2} \,
 \label{eqn:new_variables}
\end{equation}
where $\Phi_{0}$ is a constant such that the conditions $f > 0$ for
$\xi > 0$, and $f = 0$ for $\xi \leq 0$ are satisfied.

We can now derive a unique ergodic distibution function (DF) as a
function of relative energy $f(\xi)$ from the given density profile of
the sheet $\rho(|z|)$ with an infinite extension in the x- and y-
direction as
\begin{equation}
f\left ( \xi  \right ) = 
\frac{1}{\sqrt{2}\pi }  \int_{0}^{\xi }\frac{d\Psi }{\sqrt{\xi - \Psi }}
\frac{\mathrm{d} \rho }{\mathrm{d} \Psi } \, ,
\label{eqn:Eddington_1D}
\end{equation}
where $\Psi$ is the relative potential. The derivation is shown in
detail in Appendix-(\ref{Apdx1}), and this result is similar to
\citet{Eddington1916}.

We can obtain the velocity distribution function $P(v)$ along the line
of sight, by integrating $f\left ( \xi \right )$ as :
\begin{equation}
P \left ( \upsilon  \right ) = \int_{-z_{\textup{max}}}^{z_{\textup{max}}}f\left ( \xi  \right )dz \,  
\end{equation}
where $z_{max}$ is computed such that $\xi > 0$ is always valid.
$P(v)$ is directly observable, and hence this model can in principle
be used to extract the density profiles of sheets from the observation
of the line of sight velocities of the galaxies belonging to the
sheet, even though this practically may never be possible.

\section{Application to analytical models}
\label{Analytical Models}

Let us first consider a few simple and analytical models.  We will
here assume a known density profile for a sheet and calculate the
associated phase space distribution function and velocity distribution
function, $P(v)$, using equation (\ref{eqn:Eddington_1D}).  The spirit
behind this analysis is that direct observations of the sheets give us
their velocity distribution function and surface density ($\Sigma$)
only.  The detailed properties of the VDF of the sheets, obtained from
the observations, can then be compared to the computed profiles which
have the same surface density ($\Sigma$) and from it the density
profile could be deduced.  The analysis in this section is done for
idealized sheets with infinite area.

\subsection{Model 1 : An Isothermal sheet}
\label{isothermal_sheet}
First a sheet with a density profile, which is often used in the
simulations of the galaxy discs, is examined
\begin{equation}
\rho \left ( z \right ) = \begin{cases}
\frac{1}{2a}\Sigma sech^{2}\left ( \frac{z}{a} \right ) & \text{ : } \left | z \right |< z_{max}\\ 
 0& \text{ : otherwise } \, .
\end{cases}
\label{eqn:sech_density2}
\end{equation}
The scale length of the sheet is a, and the value of surface density $\Sigma$ is normalized such that $2\pi G\Sigma = 1$.

By using equation (\ref{eqn:Eddington_1D}), the phase space
distribution function is -
\begin{eqnarray}
 f\left ( \xi  \right ) 
 &=& k_{1}\int_{0}^{\xi }\frac{1}{\sqrt{\xi -\Psi }} e^{-\frac{2}{a}\left ( \Psi _{0} - \Psi \right )} d\Psi \\
 &=& k_{2}  e^{\frac{2}{a}\xi } erf\left ( \sqrt{\frac{2}{a} \xi} \right ) \, .
 \label{eqn:sech_f(e)}
\end{eqnarray}

Here `erf' represents the error function and $k_{1}$,$k_{2}$ are
normalization constants. The derivation and the approximations
involved in calculating $\Sigma$ are given in the
Appendix-(\ref{Apdx2}). In order to derive the velocity distribution
function.

\begin{equation}
P \left ( v  \right ) = k' \int_{-z_{max}}^{z_{max}}e^{\frac{2}{a}\left ( \Psi - \frac{v^{2}}{2} \right )} \times  \textup{erf}\Bigg(\sqrt{{\frac{2}{a} ( \Psi - \frac{v^{2}}{2} )}}\Bigg) dz
 \label{eqn:sech_P(v)}
\end{equation}
where $k'$ is the constant for normalization, we substitute

\begin{equation}
\Psi = \Psi _{0} - a   \ln\left ( \frac{e^{\frac{z}{a}} + e^{-\frac{z}{a}}}{2} \right ) \,  . 
\end{equation}

\begin{figure}
\centering
\includegraphics[scale=0.7,keepaspectratio=true]{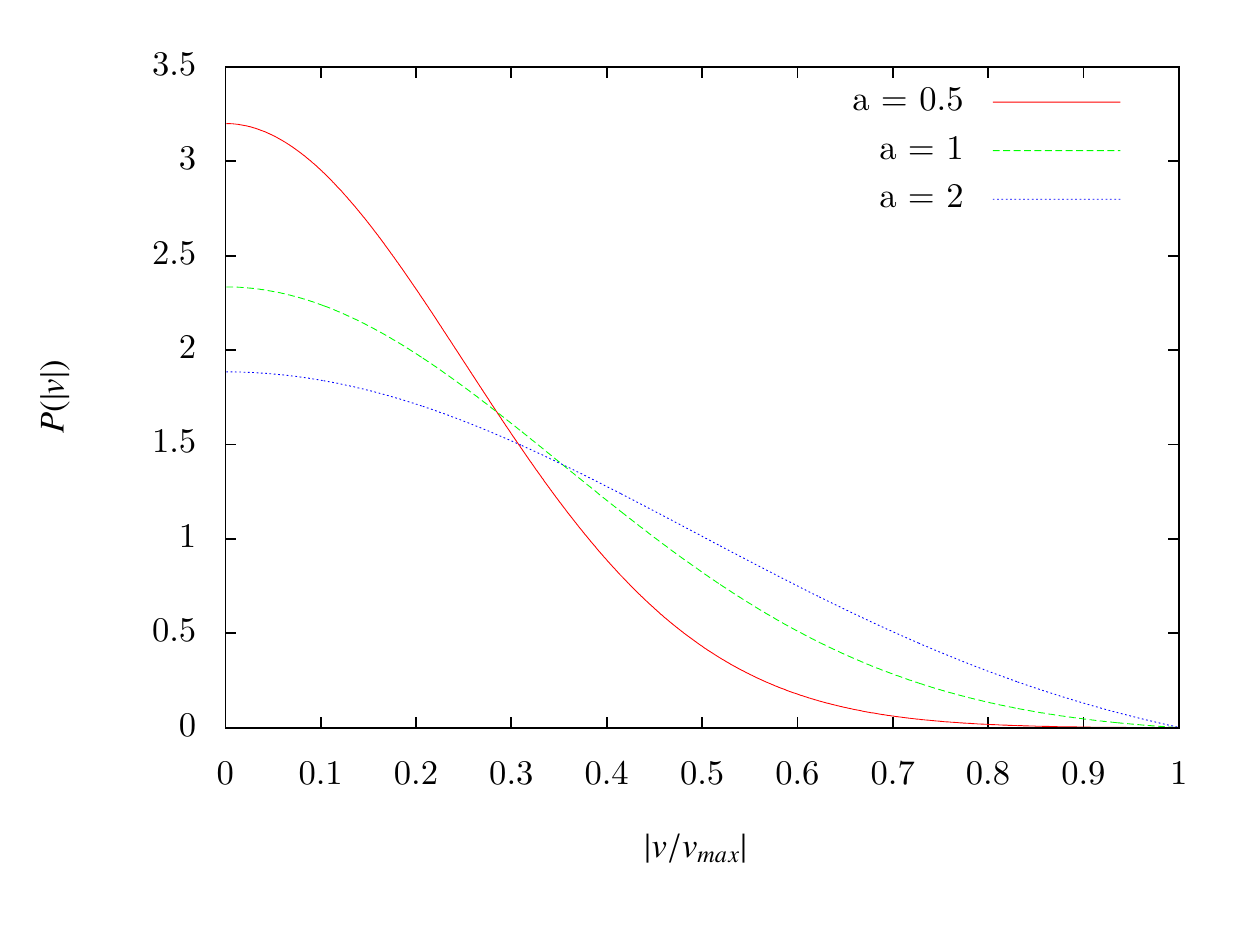}                                                                   
\caption{The velocity distribution functions for density
  profile represented by equation (\ref{eqn:sech_density2}) for
  different values of constant $a$.  The value of mass per unit area
  $(\Sigma)$ is kept constant for all the curves.}
\label{fig:Sech_VDF}
\end{figure}

The definite integral can be computed to get the distribution of
velocities.  The resulting profile obtained is shown in figure
(\ref{fig:Sech_VDF}) for different values of $a$.

\par
\subsection{Model 2 : A core with a power law cutoff}
Another density profile with convenient analytical properties is
\begin{equation}
  \rho \left ( z \right ) = \begin{cases}
    \Sigma \frac{a^{2}}{2}\left ( z^{2} + a^{2} \right )^{-\frac{3}{2}} & \text{ : } \left | z \right |< z_{max}\\ 
    0& \text{ : otherwise }  \, .
\end{cases}
\label{eqn:Power_Law}
\end{equation}
The scale length of the sheet is $a$ and it's surface density is
($\Sigma$). The equation for the relative potential becomes
\begin{equation}
 \Psi_{0} - \Psi = 2\pi G \Sigma (\sqrt{z^{2}+a^{2}} - a)
\end{equation}

Similar to the previous example, the value of surface density
($\Sigma$) is considered such that $2\pi G\Sigma = 1$.  The phase
space distributions function for two values of $a$ are given below.

\subsubsection{\textbf{a = 1}}

\begin{eqnarray*}
  f(\xi ) &=& \frac{\sqrt{\xi }}{648(3-\xi )^{3}}\Big(8\xi^{2} - 78\xi +297 \Big) +\\
  && + \frac{5}{8(3-\xi )^{\frac{7}{2}}}\tan^{-1} \Big(\sqrt{\frac{\xi }{3-\xi }}\Big)
\end{eqnarray*}

\subsubsection{\textbf{a = 2}}

\begin{eqnarray*}
  f(\xi ) &=& \frac{\sqrt{\xi }}{1536(4-\xi )^{3}}\Big(8\xi^{2} - 104\xi +528 \Big) +\\
  && + \frac{5}{8(4-\xi )^{\frac{7}{2}}}\tan^{-1} \Big(\sqrt{\frac{\xi }{4-\xi }}\Big)
\end{eqnarray*}

\begin{figure}
\centering
\includegraphics[scale=0.7,keepaspectratio=true]{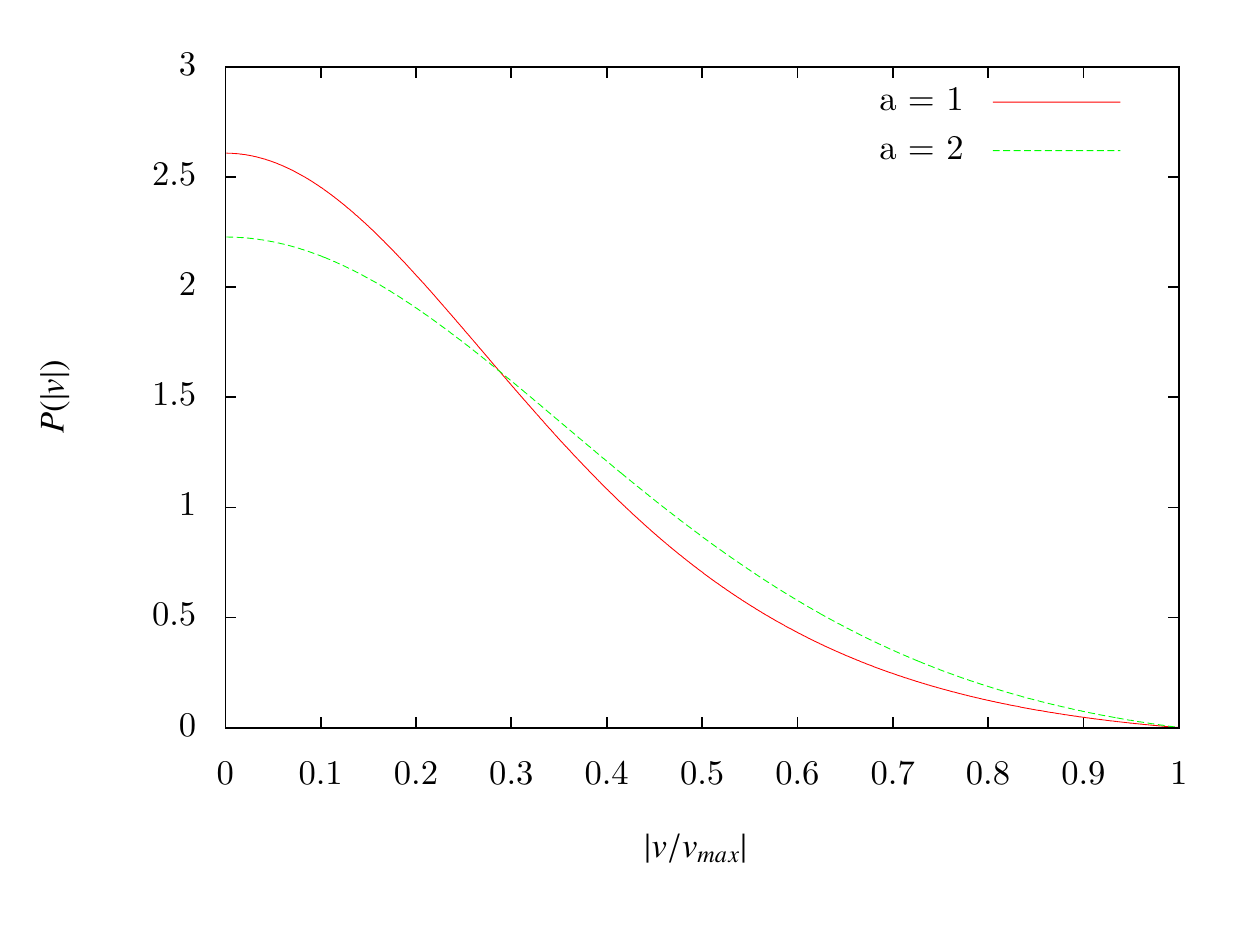}                              
\caption{The velocity distribution
  functions for density profile represented by equation
  (\ref{eqn:Power_Law}) for different values of constant $a$.  The
  value of mass per unit area $(\Sigma)$ is kept constant for all the
  curves.}
\label{fig:Power_Law_VDF}
\end{figure}

Further, the velocity distribution profile can be obtained by direct
integration and resulting profiles obtained are shown in figure
(\ref{fig:Power_Law_VDF}).

\section{Application to simulations}

\begin{figure*}
\vbox to 80mm {
\hbox {\includegraphics[scale=1.0,keepaspectratio=true]{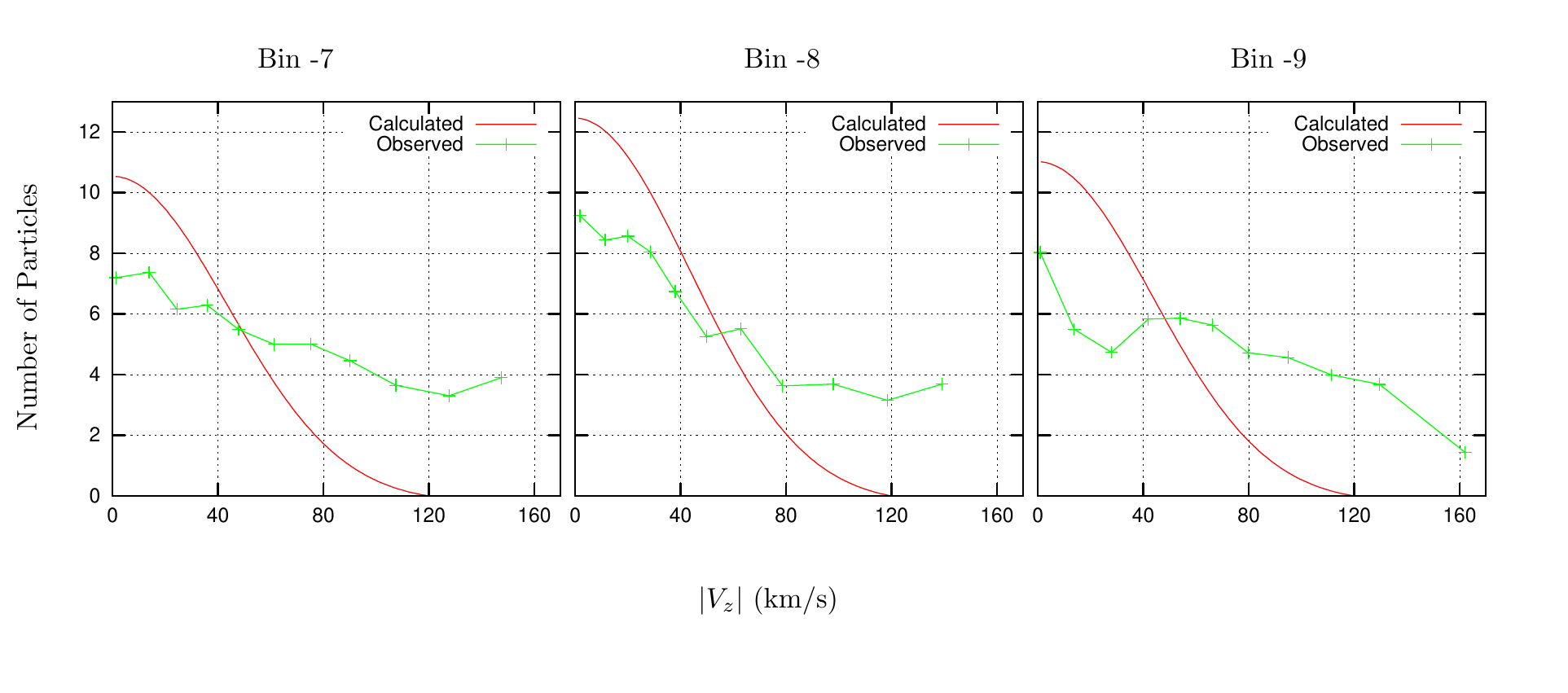}}
}

\vspace{0.5cm}
\caption{This figure shows the comparison between the observed
  velocity distribution curves (in green) and the curves derived from
  the application of Eddington model (in red).
  The observed velocity distribution curve falls
  less steeply than the calculated one which implies that the sheet
  isn't completely equilibrated yet. }

\label{fig:Application-Eddington}
\end{figure*}

In this section we will present a comparison between the derived
Eddington Model given by the equation (\ref{eqn:Eddington_1D}) on the
simulated CDM sheet described earlier section. We will apply the
Eddington formula separately on the 7th, 8th \& 9th radial bins of the
largest sheet and compare to the analytically calculated velocity
distribution curves with the observed VDFs for each of the bins.  The
calculation is quite tedious due to various approximations and
complicated functions are involved in it.

First of all, for the density distribution
 along the perpendicular axis of the sheet, we will estimate the best-fit parameters 
for the symmetric analytical model presented in section (\ref{isothermal_sheet}).
In this step, we have ignored the anisotropies present in the sheet along its perpendicular direction (Z-axis).
The gravitational potential
function of the sheet is then calculated from the density
distribution using the Poisson's
equation (\ref{eqn:Poisson_Equation}). 
In this analysis, we are dealing with particles with z coordinates less than 
the dimensions of the sheet in the xy plane so we can perform the potential energy calculations
similar to the simplified calculation for a sheet of infinite area.

For Eddington formula to be applicable, a structure must be completely
equilibrated so that the ergodic hypothesis is valid.  The radial bins
of the sheet are divided into structures as discussed in section
\ref{Phase_space_structure}.  

To make a proper comparison between
calculated and observed VDFs, we will take into consideration only the
particles which lie in equilibrated regions of the sheet for plotting
the observed VDF and we will neglect the particles lying in the
infalling streams. Using these particles, we can also fix an
appropriate value to the width ($z_{max}$) of the sheet which is an
input parameter needed for performing the integration in equation
(\ref{eqn:sech_P(v)}).

The computed and observed
total velocity distribution curves for each of the bins 7,8,9 are
compared separately in the figure (\ref{fig:Application-Eddington}).
The constraints on $z _{max}$ can be used to roughly calculate $v_{max}$
which turns out to be almost $120 \, km/s$ in these cases.
These curves can be observed to be similar within an order of magnitude.
The discrepancies between the calculated and observed VDFs can be attributed 
to the various approximations which are elaborated in this section.

\section{Conclusions}
Zeldovich pancakes are cosmological structures which have collapsed
and equilibrated only along the one dimension.  These pancakes can be
identified at redshift zero using only observational parameters and
have recently received renewed interest, since they can be used to
measure cluster masses.

We consider the phase-space properties of pancakes from a cosmological
numerical simulation, which were detected using the observational
technique.  We demonstrate how these objects are in the process of
equilibrating mainly along the one dimension, and are much less
equilibrated along the other two. We extract the density profiles of
these structures, and we find fair agreement with theoretical
predictions.

\section*{Acknowledgments}
It is a pleasure thanking Martina Falco, Radek Wojtak, Martin Sparre, Thejs Brinckmann and 
Mikkel Lindholmer for discussions.
The Dark Cosmology Centre is funded by the Danish National Research Foundation.

\def\aj{AJ}
\def\araa{ARA\&A}
\def\apj{ApJ}
\def\apjl{ApJ}
\def\apjs{ApJS}
\def\apss{Ap\&SS}
\def\aap{A\&A}
\def\aapr{A\&A~Rev.}
\def\aaps{A\&AS}
\def\mnras{MNRAS}
\def\jcap{JCAP}
\def\nat{Nature}
\def\pasp{PASP}
\def\aplett{Astrophys.~Lett.}
\def\physrep{Physical Reviews}
\def\prd{ Phys. Rev. D}
\def\rvmp{RvMP}
\def\gapfd{GApFD}
\def\comap{Comments on Astrophysics}

\bibliographystyle{mn2e}

\bsp

\appendix

\section{Derivation of one dimensional Eddington Model}
\label{Apdx1}

This section shows the derivation of an equivalent of Eddington Model
for galactic sheets which was earlier presented in
equation~(\ref{eqn:sech_P(v)}).  These isotropic sheets have infinite
extension in the x- and y- direction, and are completely equilibrated
along their perpendicular dimension with a known density profile
$\rho \left ( \left | z \right | \right )$.  The Eddington model was
originally derived for completely equilibrated three dimensional
structures \citep{Eddington1916, BinneyBook}.

We have defined the relative energy and relative potential in
equation(\ref{eqn:new_variables}).  The relative potential for an
isolated system can be calculated using Poisson's equation

\begin{equation}
\Delta \Psi =  -4\pi G\rho \, .
\label{eqn:Poisson_Equation}
\end{equation}

It is possible to derive for the
system a unique ergodic distribution function $(f)$ that depends upon the
phase-space coordinates only through the Hamiltonian $H\left ( z,v
\right )$ so we can write the DF as a function of the relative energy $f\left (
  \xi \right )$. The number density  $\rho \left (
  \left | z \right | \right )$ is an integral of $f\left (
  \xi \right )$ over all velocities. The sheet is completely isotropic thus,

\begin{subequations}

\begin{align}
\rho \left ( \left | z \right | \right ) 
&= 2\int d\upsilon  f\left ( \Psi - \frac{\upsilon ^{2}}{2} \right )\\
&= 2\int_{0}^{\Psi }d\xi \frac{f\left ( \xi  \right )}{\sqrt{2\left ( \Psi - \xi  \right )}} \, .
\end{align}

\end{subequations}
Here we have used the definitions in equation (\ref{eqn:new_variables}) and the value of constant $\Phi_{0}$ in $\xi$ 
is chosen such that $\Phi = 0$ and  $\xi = 0$
at the boundary of the sheet and $f = 0$ for $\xi \leq 0$. We can write $\rho$ as a function of $\Phi$
instead of $ |z| $ since $\Phi$ is a monotonic function of $ |z| $.

\begin{equation}
\frac{\rho \left ( \Psi  \right )}{\sqrt{2}} 
= \int_{0}^{\Psi } d\xi \frac{f\left ( \xi  \right )}{\sqrt{\Psi - \xi }} \, .
\end{equation}

Differentiating both sides with respect to $\Phi$, we obtain
\begin{equation}
\frac{1}{\sqrt{2}}\frac{\mathrm{d}\rho  }{\mathrm{d} \Psi } = 
\frac{\mathrm{d} }{\mathrm{d} \Psi } \int_{0}^{\Psi } d\xi  
\frac{f\left ( \xi  \right )}{\sqrt{\left ( \Psi - \xi  \right )^{}}} \, .
\end{equation}

Note that the integral on the right is the convolution of $f\left ( \xi  \right )$
with \large {$1/\sqrt{\xi }$}.
Thus taking Laplace transform of both sides
\begin{equation}
\frac{1}{\sqrt{2}} L  \left [\frac{\mathrm{d} \rho }{\mathrm{d} \Psi }  \right ]  
= s\times L \left [f\left ( \xi  \right )  \right ] \times  L\left [ \frac{1}{\sqrt{\xi }} \right ] \, .
\end{equation}
Since, 
\begin{equation}
L\left [ \frac{1}{\sqrt{x}} \right ] = \sqrt{\pi }s^{-\frac{1}{2}} \, ,
\end{equation}
and on further simplification,
we now have an expression for the Laplace transform of $f\left ( \xi  \right )$
in terms of $L  \left [\frac{\mathrm{d} \rho }{\mathrm{d} \Psi }  \right ]$ as:

\begin{equation}
L\left [ f\left ( \xi  \right ) \right ] 
= \frac{1}{\sqrt{2 }\pi } \left (\sqrt{\pi }s^{-\frac{1}{2}}  \right )
\times L  \left [\frac{\mathrm{d} \rho }{\mathrm{d} \Psi }  \right ]
\end{equation}

Applying the inverted laplace transform on both sides we conclude

\begin{equation}
f\left ( \xi  \right ) = 
\frac{1}{\sqrt{2}\pi }  \int_{0}^{\xi }\frac{d\Psi }{\sqrt{\xi - \Psi }}
\frac{\mathrm{d} \rho }{\mathrm{d} \Psi } \, .
\label{eqn:Eddington}
\end{equation}
This result is similar to \citep{Eddington1916}.  It also
follows from the equation (\ref{eqn:Eddington}), that the density
distribution for the sheet $v(z)$ in the potential $\Phi(z)$ can arise
from an ergodic DF  if and only if
\begin{equation}
 \int_{0}^{\xi }\frac{d\Psi }{\sqrt{\xi - \Psi }}
\frac{\mathrm{d} \rho }{\mathrm{d} \Psi } > 0 \, .
 \end{equation}

Further to get the velocity profile we need to integrate $f\left ( \xi  \right )$ as :
\begin{equation}
 P \left ( \upsilon  \right ) = k \int_{-z_{\textup{max}}}^{z_{\textup{max}}}f\left ( \xi  \right )dz
\end{equation}

where k is the constant of integration. The range of allowed velocities to keep $\xi$ positive 

\begin{equation}
 -\sqrt{2\Psi _{0}} < v < \sqrt{2\Psi _{0}} \, .
\end{equation}

\section{Derivation of the P(v) curve for an Isothermal sheet}
\label{Apdx2}
This section shows the derivation of velocity distribution function for an isothermal sheet
 which was earlier presented in equation(\ref{eqn:Eddington_1D}). 
We had used a sheet with a density profile as

\begin{equation}
\rho \left ( z \right ) = \begin{cases}
\frac{1}{2a}\Sigma sech^{2}\left ( \frac{z}{a} \right ) & \text{ : } \left | z \right |< z_{max}\\ 
 0& \text{ : otherwise }  \, ,
\end{cases}
\label{eqn:sech_density}
\end{equation}
where the scale length of the sheet is a, and the surface density is defined as 
\begin{equation}
\Sigma = \int_{-z_{\textup{max}} }^{z_{\textup{max}}}\rho \left ( z \right )dz  \sim \int_{-\infty }^{\infty }\rho \left ( z \right )dz \, .
\end{equation}

We have used the above approximation considering that $z_{max}$ is
large enough otherwise the value of $\Sigma$ would depend on value of ($a$) to keep the value of surface density constant.
In the analytical models for sheets the potential along Z axis increases more steeply with distance
so to keep the value of $\Psi _{0}$ finite we need to constrain $Z_{max}$ to an appropriate value.
The relative potential can be calculated as

\begin{equation}
\Psi = \Psi _{0} - 2a\pi G\Sigma   ln\left ( \frac{e^{\frac{z}{a}} + e^{-\frac{z}{a}}}{2} \right ) \, .
\label{eqn:sech_Psi}
\end{equation}

We will consider a sheet with the value of $\Sigma$ normalized such that $2\pi G\Sigma = 1$ to reduce the number of constants. When we use the constraint for 
$f\left ( \xi  \right )$ being positive, the value of $\Psi_{0}$ for $z=0, v=0$ is 
\begin{equation}
 \Psi _{0} = \Phi _{0} - [\Phi (= 0)] = a  \ln \left ( \frac{1}{2}(e^{\frac{z_{\textup{max}}}{a}} + e^{-\frac{z_{\textup{max}}}{a}}) \right ) \,.
\end{equation}

Inverting the coordinate `z' as a function of $\Psi$ we get,

\begin{equation}
z = a\ln\left ( e^{\frac{1}{a}(\Psi _{0}-\Psi )} + \sqrt{e^{\frac{2}{a} (\Psi _{0}-\Psi )}  - 1} \right )
\end{equation}

Using the expressions for $\rho$ \& $\Psi$

\begin{equation}
 \frac{\mathrm{d} \rho }{\mathrm{d} \Psi } = \frac{1}{2a^{2} \pi G} sech^{2}\left ( \frac{z}{a} \right )
\end{equation}

Using the expression in equation (\ref{eqn:sech_Psi}) we can use the formula(\ref{eqn:Eddington})
to obtain

\begin{eqnarray}
 f\left ( \xi  \right ) 
 &=& k_{1}\int_{0}^{\xi }\frac{1}{\sqrt{\xi -\Psi }} e^{-\frac{2}{a}\left ( \Psi _{0} - \Psi \right )} d\Psi \\
 &=& k_{2}  e^{\frac{2}{a}\xi } \textup{erf}\left ( \sqrt{\frac{2}{a} \xi} \right )
\end{eqnarray}

Here `$erf$' represents the error function and $k_{1}$,$k_{2}$ are
normalization constants. Further we substitute back the formula for
relative energy from equation (\ref{eqn:new_variables}) to derive the
velocity distribution function as

\begin{equation}
  P \left ( v  \right ) = k' \int_{-z_{\textup{max}}}^{z_{\textup{max}}}e^{\frac{2}{a}\left ( \Psi - \frac{v^{2}}{2} \right )} \times  \textup{erf}\Bigg(\sqrt{{\frac{2}{a} ( \Psi - \frac{v^{2}}{2} )}}\Bigg) dz
\end{equation}
where $k'$ is the constant for normalization and we can substitute
$\Psi$ from equation (\ref{eqn:sech_Psi}).  This definite integral is
too tedious to compute analytically and can be computed numerically
when the values of all variables are provided.

\label{lastpage}

\end{document}